\documentstyle[eqsecnum,floats,prd,aps]{revtex}

\begin{document}
\def\vp{\varphi}
\def\square{\kern1pt\vbox{\hrule height 1.2pt\hbox{\vrule width 1.2pt\hskip 3pt
   \vbox{\vskip 6pt}\hskip 3pt\vrule width 0.6pt}\hrule height 0.6pt}\kern1pt}

\draft

\title{Curing Singularities:   From the Big Bang to Black Holes}
\author{Janna Levin}
\address{Center for Particle Astrophysics,
UC Berkeley, Berkeley, CA 94720-7304}

\twocolumn{
\maketitle
\widetext
\begin{abstract}

Singular spacetimes are a natural prediction of Einstein's theory.
Most memorable are the singular centers of black holes and the big
bang.  However, 
dilatonic extensions of Einstein's theory can support nonsingular
spacetimes.  The cosmological singularities can be avoided by dilaton driven
inflation.  Furthermore,
a nonsingular black hole can be constructed in two dimensions.

\end{abstract}
\pacs{}
}
\twocolumn
\narrowtext
\begin{picture}(0,0)
\put(410,190){{ CfPA-97-TH-08}}
\end{picture} \vspace*{-0.15 in}
\setcounter{section}{1}

The big 
bang and black holes are both spectacular predictions of Einstein's theory
which share a singular nature.
The singularities mark the breakdown of the classical theory,
as predictability of the future or past is lost for world-lines
which run into or out of the singularity.  
Infinite energy scales are invoked as the
curvature invariants become infinite.
The ultimate theory of quantum gravity, it is often hoped, will
temper these singularities.  
Still, even without the full quantum theory, 
singularities 
can be avoided classically.  

As argued by the Hawking-Penrose
theorems,
Einstein gravity plus any ordinary
matter will spawn singular spacetimes
\cite{hp}.
The singularity theorems assume that both the Einstein equations 
hold and that the matter sector obeys the strong energy condition.
For a diagonal 
energy-momentum tensor $T^\mu_\nu=(-\rho,p_1,p_2,p_3)$,
the weak energy condition states 
	\begin{equation}
	\rho\ge 0\quad \quad {\rm and}\quad \quad \rho +p_{1,2,3}\ge 0
	\ \ ,
	\end{equation}
while the strong energy condition states
	\begin{equation}
	\rho+p_1+p_2+p_3\ge 0\quad \quad {\rm and}\quad \quad 
	\rho+{p_{1,2,3}}\ge 0
	\ \  .
	\end{equation}
(See for instance \cite{wald}.)
With the advent of inflation, violations of the strong energy condition 
are commonplace.
The negative pressure needed to drive the inflationary growth of the
universe can produce $p\le -\rho $, 
though it is still advisable to respect the positivity 
of the energy density, $\rho > 0$.
If the energy conditions are violated, the singularity theorems
do not hold.  
While this
does not ensure a
nonsingular universe,
many inflationary cosmologies can 
be shown to be nonsingular.

In general,
simple conditions can be found on the matter stress tensor 
such that the minisuperspace of all homogeneous, isotropic cosmologies
($ds^2=-dt^2+a^2d\ell^2$)
is nonsingular.   For this subset of all possible metrics, the curvature
invariants depend only on $H,\dot H$ and $a$ where $H=\dot a/a$.
An {\it initial} singularity will be avoided for an expanding universe if
	\begin{equation}
	\dot H =-{4\pi \over M_{PL}^2}\left(\rho+p\right )\ge 0
	\ \ ;\label{c1}
	\end{equation}
that is, if $p\le -\rho$ and the energy 
conditions on which the singularity theorems are based
are violated.  
Eqn (\ref{c1}) also ensures that null geodesics
do not converge, $R_{\mu \nu}n^\mu n^\nu \propto -\dot H \le 0$
indicating geodesic completeness \cite{bv}.
While eqn (\ref{c1}) skirts an initial singularity, there may still be a {\it
future} singularity, particularly if the model is superinflationary,
i.e., $p<-\rho$ and so $\dot H>0$.  To avoid the future singularity,
the evolution should roll over to 
	\begin{equation}
	{\rm sgn}(H)\dot H
	=-{\rm sgn}(H){4\pi \over M_{PL}^2}\left(\rho+p\right )\le 0
	\ \  \label{c2}
	\end{equation}
before the singularity strikes.
The sign of $H$ incorporates the possibiity
that the universe contracts.
For an initially expanding universe, 
condition (\ref{c2}) requires $H$ grow faster than or equal
to a constant.  As we look back in time, $H$ decreases and therefore
must have been less than infinite.  Similarly $a$ grows faster than or
equal to $e^t$ and so is finite as $t\rightarrow 0$.  If we push time
back to $-\infty$, then the space can be continued onto an initially
large and contracting universe.  Condition (\ref{c2}) then requires
that $\dot H<0$.  In the past, $H$ gets less negative and so 
must have been finite.  

It may still be possible to find spacetimes which meet the above criteria
but are singular in higher derivatives of $H$ \cite{priv}.  
Protection against singularities in higher
derivatives of $H$ requires ${\rm sgn}(d^n H/dt^n)d^{n+1}H/dt^{n+1}$
positive in the past and negative in the future.
The
curvature invariants depend also on $\dot H$.  
If they are to be finite, 
the equation of state for matter is restricted by
$|p |<\infty \ \rho$, a reasonable requirement.  In short, if matter
obeys an equation of state bounded by
$-\infty < p < 0$ in the past and $0 < p < \infty $ in the future, 
then the spacetime should be nonsingular.

The maximally symmetric de Sitter cosmology is the simplest nonsingular,
inflationary universe.  The energy momentum tensor is provided by 
a cosmological constant $\Lambda $ with $\rho_\Lambda=- p_\Lambda$.  
In standard inflationary models these nonsingular
conditions are difficult to 
maintain \cite{bv}.
All energy densities which scale as $\rho\sim a^{-\alpha} $  will
dominate over any constant cosmological density as 
the scale factor goes to zero,  
thereby reinstating a singularity.  
A generic prescription for avoiding precisely this problem
has been based on de Sitter regularity \cite{brand}.
The prescription is for a limiting curvature with a kind of asymptotic freedom.
The coupling between gravity and matter becomes weaker as the 
limiting curvature is reached and de Sitter evolution reigns.
Superinflationary models may be more robust than pure de Sitter.
For these, the inflationary component becomes more effective as 
the would be singularity is approached.

There are very few forms of matter which induce negative pressures.
Again, there is the 
celebrated cosmological constant.
Less well known, a gas of fundamental strings can have negative pressure
\cite{turok}.
Before the advent of the now standard inflationary
cosmology,
Starobinsky's ${\cal R}^2$ inflation
was first advanced as
a nonsingular big bang \cite{staro}.  
Another modification to Einstein's theory,
motivated both by ordinary quantum corrections and string theories,
is generalized dilaton-gravity.  Classes of such models have been found
to generate a negative pressure \cite{{me},{ven}}.
Recently nonsingular dilaton cosmologies were selected on the basis
of their kinetic coupling \cite{rama}.  
It can be shown that all of these kinetic
couplings lead to negative pressures and kinetic inflation.
These models are ideal for satisfying the conditions 
(\ref{c1})-(\ref{c2})
since they can begin superinflationary and then connect onto
a regular decelerating cosmology \cite{me}.

To illustrate how the dilaton avoids singularities consider the
theory of gravity
	\begin{equation}
	A=\int d^4x\sqrt{-g}
	\left [-\vp {\cal R}+{\omega\over \vp}(\partial \vp)^2
	\right ] \ \ .
	\label{act}
	\end{equation}
The lowest energy effective action from string theory predicts $\omega=-1$.
The string theory dilaton leads to superinflation and
begins nonsingular.  In fact, the cosmos begins asymptotically flat \cite{ven}.
However, it quickly runs into a 
problematic future singularity.  In general,
higher order string contributions can generate a variable $\omega(\vp)$
\cite{dp}.  
Regardless of motiviation, $\omega(\vp)$ may be chosen such that
the universe is nonsingular in both the past 
and the future \cite{{me},{rama}}.

Consider for the sake of simplicity
$\omega=-4/3$. The evolution then appears identical to a de Sitter spacetime.
The kinetic energy in the Planck field is
	\begin{equation}
	\rho_\vp/\vp\propto \left ( {\dot \vp\over \vp}\right )^2 =
	{\rm constant}\ \ .
	\end{equation}
Then $H^2=$constant and 
$a\sim e^t$ while $\vp\sim e^{-3t}$.
All of the curvature invariants are de Sitter and nonsingular.
A simple test particle moving along the space has no means by which to distinguish
this from de Sitter.

While the curvature invatiant is finite and in this sense the space is 
nonsinuglar, there ate still some subtleties.
All models of the form (\ref{act}) are degnerately conformal to a singular
universe described by Einstein gravity plus a 
minimally coupled scalar field.  However, there
is no paradox since the conformal transformation 
connecting the two is singular.
Thus the singular Einstein space is only conformal to a portion of the
nonsingular dilaton space.
An analogous 
relationship arises with the common tool of
transforming from expanding to comoving
coordinates.
A singular expanding space is conformally transformed to 
a nonsingular, comoving Minkowskii space 
such that the conformal transformation carries
the singularity.
(Having said this however, 
gravity wave probes may be sensitive to the behaviour
as $t\rightarrow -\infty$ as argued in Ref.
\cite{kal}.  If this is so, the space while bearing finite curvature
invariants is not stably nonsingular.)

The limiting curvature hypothesis was used in $(1+1)$ dimensional 
gravity to locate nonsingular $2D$ black holes \cite{brand}.
A potential was chosen to support the dilaton and the 
spacetime in a nonsingular configuration.
\footnote{
The potential chosen in Ref. 
\cite{brand} may actually permit a singularity if $\vp$
is allowed negative values.}
Since those solutions are also dilaton motivated it is of interest
to show how a choice of the kinetic coupling can accomplish the same
task.  A dilaton-metric configuration
similar to that of Ref.\cite{brand} is found below which is supported by an
astute choice of the kinetic coupling.

The action is the two dimensional version of (\ref{act}).
A metric of the form 
	\begin{equation}
	ds^2=-h(r) dt^2+h^{-1}(r)dr^2
	\end{equation}
is sought.
The trace of $G_{\mu \nu}=T_{\mu \nu}=0$ leads to 
	\begin{equation}
	\square \Phi=0 \quad\quad \Rightarrow \quad \quad \vp^{\prime \prime}=0
	\ \ .
	\end{equation}
The constants of integration are chosen such that
$\vp=r$.
\
\begin{figure}[h]
\vspace{56mm}
\includegraphics{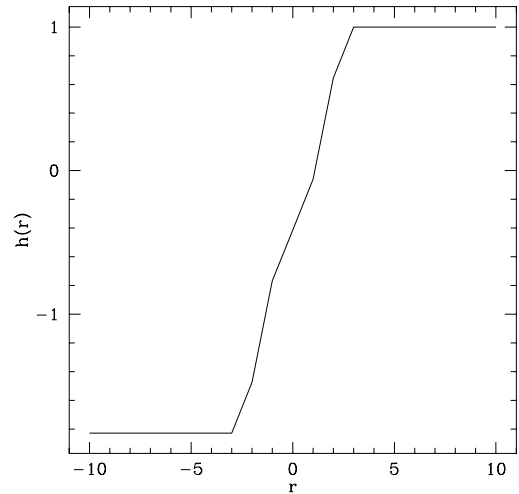}
\vspace{15mm}
\caption{The metric component $h(r)$.
\label{ah}} 
\end{figure} 

The Einstein Equations require
	\begin{equation}
	h^\prime={\omega\over \vp} \ \ ,
	\label{hpr}
	\end{equation}
and the only curvature invariant is
	\begin{equation}
	{\cal R }= 
	\left (\omega\over \vp\right )^{\prime} =h(r)^{\prime \prime}
	\ \ .
	\end{equation}
The behaviour desired is for the metric to behave
as a normal Schwarzschild solution
far from the horizon;
	\begin{equation}
	\lim_{r\rightarrow \infty} \quad h(r) \rightarrow_{} 
	\quad \left (1-{2m\over r}\right )
	\ \ .\label{metric}
	\end{equation}
Nonsingularity requires
	\begin{equation}
	\lim_{r\rightarrow 0}	\quad {\cal R} \rightarrow_{} \quad 
	{\rm finite}.\label{fin}
	\end{equation}
An example of a  kinetic coupling, $\omega$, which satisfies both 
(\ref{metric}) and (\ref{fin}) is
	\begin{equation}
	{\omega\over \vp}={2 m \vp^2\over \vp^4+m^4} \  \ .
	\label{om}
	\end{equation}
Eqn (\ref{hpr}) can be integrated and is depicted in Fig. \ref{ah}.
Notice, it passes through zero but is never infinite.
There is a Killing vector $\xi=(h,0)$
which becomes null at $h(r)=0$ signaling the occurence of an 
event horizon.
Consequently, photons emitted from that point are infinitely redshifted
and the surface is black.

\begin{figure}[h]
\vspace{56mm}
\includegraphics{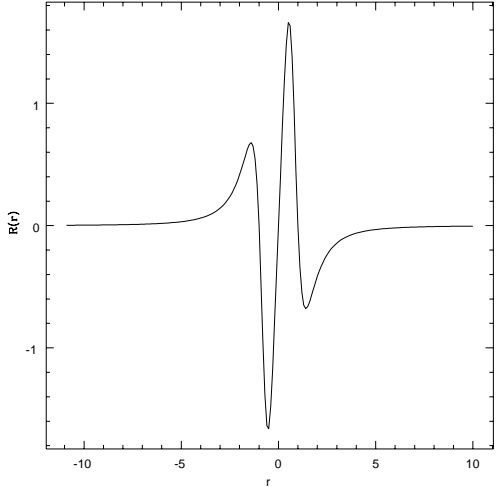}  
\vspace{15mm}
\caption{The nonsingular Ricci scalar.\label{ricci}}  \end{figure}

The curvature invariant
	\begin{equation}
	{\cal R}=-{4mr(r^4-m^4)\over (r^4+m^4)^2}
	\end{equation}
is zero as $r\rightarrow \infty$ and
zero as $r\rightarrow 0$. 
It is nowhere singular as shown in Fig. \ref{ricci}.
\

This spacetime appears geodesically complete also
as can be seen with the Lagrangian,
\begin{equation}
	{\cal L}={1\over 2}h\dot t^2-{1\over 2}h^{-1}\dot r^2
	\end{equation}
where
an overdot denotes differentiation with respect to an affine
parameter $\lambda $.
The metric is static so that 
$h\dot t=E$, which is chosen to be $E\equiv 1$.
Null geodesics lie on $h\dot t^2-h^{-1}\dot r^2=0$ so that
$\dot r^2=1$.
Combine the two geodesic equations to find
	\begin{equation}
	\left ({dr\over dt}\right )^2=h^2(r) \ \ .
	\end{equation}
Notice that 
	\begin{equation}
	{dr^2\over dt^2}=- h^{\prime}h \quad \lim_{r \rightarrow 0}\quad 
	\rightarrow_{  }\quad 0 .
	\end{equation} 
There is no force at the center, unlike a singular black hole for 
which the acceleration is infinitely negative driving the photon 
out the  cut in spacetime.

\begin{figure}[h]
\vspace{30mm}
\includegraphics{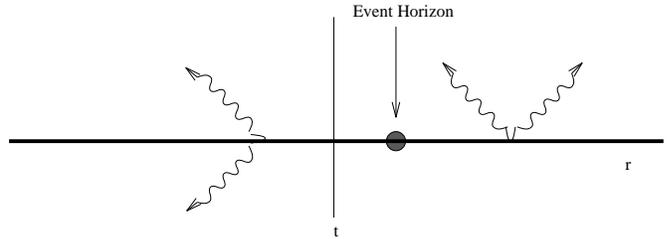}  
\vspace{15mm}
\caption{A schematic spacetime diagram of the black hole or rather, the black
point.  Photons emitted from the point indicated are infinitely 
redshift to the right.  Light which is drawn through the event
horizon moves forever to the left as represented by the tilting of the
light cones.\label{eh}}  \end{figure}

At large $r$ this has the features of a black hole.  At some finite
$r$ there is a surface of infinite redshift and small curvature.
Inside, the curvature rises to 
some maximum value but then turns over and vanishes at the center.
The event horizon is only one point as shown in Fig. \ref{eh}.  Once inside the
event horizon, photons move forever along the axis to an asymptotically
flat space.  It is sensible that there are no truly trapped surfaces.
If there were, photons would have nowhere to go but out a singularity.
For this reason, 
a $4D$ black hole would have to grow an infinitely long throat
or perhaps even a nested expanding universe \cite{rpriv}.

\

While this has all of the features of a black hole there is still
some ambiguity in the interpretation of the spacetime.
In particular, in $2D$, the space can be conformally mapped
by $g_{\mu \nu}=\exp{\left(-4T(\vp)\right )}g_{\mu \nu}^T$
onto an everywhere flat spacetime with the action
	\begin{equation}
	A=\int d^2x\sqrt{-g_T}\left [-\vp {\cal R}_T\right ]
	\ \  .
	\end{equation}
There is no kinetic coupling to choose.
Variation with respect to $\vp$ gives
${\cal R}_T=0$ everywhere.
The transformation 
	\begin{equation} 
	T(\vp)=2\int {dr\over h^{\prime}} ={r^3\over 3m}-{m^3\over r}
	\ \ ,
	\end{equation}
is singular both at $0$ and $\infty$.  However, any observer who measures the
wavelength of a photon emitted from near the black surface will perceive an
infinite redshift. Observers believe they live in a flat space but that their
rulers are warped throughout space by the coupling to the field $\vp$.
When the 
observer measures the wavelength of a photon,
the ratio of the wavelength to the ruler length is a conformal
invariant.
In this sense, the same point appears dark in both frames.


Classical dynamics can render both
a universe and $2D$ black holes nonsingular.
The interiors of nonsingular black holes are even able to support
inflation, thus the destructive singularity is avoided in favor of 
creating a universe \cite{{mm},{brand}}.
Not only can singularities be
avoided but so too can all quantum gravity scales. 
While nonsingular solutions are noteworthy, it is fair to say that there is as
yet nothing natural about these singularity avoidance mechanisms.
Further they may not be stable
\cite{{bv},{kal}}.  To put it another way, they may not
be attractors in the space of all possible solutions.
Still, the possibility of a nonsingular cosmos offers an interesting
sketch of the early universe.  The violent energetic beginning is
replaced with a smooth classical state.
Creation of such a nonsingular universe from nothing may be 
less costly.
Inflation then transports the cosmos into a big bang epoch which culminates
in a universe vast and energetic enough for us to
witness.

\vskip 10truept

I am grateful to J.R. Bond, A. Borde, R. Brandenberger, P. Ferreira,
N. Kaloper, 
and K. Rama for 
valuable contributions to these ideas.
This research is supported in part by
a President's Postdoctoral Fellowship.


\begin{references}
\bibitem{priv} One such example was pointed out to me in a private
communication with K. Rama.
\end{references}
\end{document}